\documentclass[twocolumn,aps,prapplied,longbibliography,showpacs,superscriptaddress]{revtex4-2}
\usepackage{graphicx}
\usepackage[OT1]{fontenc}
\usepackage[usenames,dvipsnames]{pstricks}
\usepackage{lmodern} 
\newcommand{\mycol}{1}
\newcommand{\beq}{\begin{equation}}
\newcommand{\eeq}{\end{equation}}
\usepackage{amsmath}
\usepackage{stackengine}
\usepackage{color}
\usepackage{gensymb}
\usepackage{float}
\usepackage{hyperref}
\usepackage{subfigure}
\usepackage{tabularx}
\usepackage{comment}
\usepackage{}
\begin{document}

\title{On-Chip Detection of Electronuclear Transitions in the $^{155,157}$Gd Multilevel Spin System}

\author{G.~Franco-Rivera}
\email{gfranco@magnet.fsu.edu}
\affiliation{Department of Physics and The National High Magnetic Field Laboratory, Florida State University, Tallahassee, Florida 32310, USA}

\author{J.~Cochran}
\affiliation{Department of Physics and The National High Magnetic Field Laboratory, Florida State University, Tallahassee, Florida 32310, USA}

\author{L.~Chen}
\affiliation{Center for Excellence in Superconducting Electronics, Shanghai Institute of Microsystems and Information Technology, Chinese Academy of Science, Shanghai, China.}

\author{S.~Bertaina}
\affiliation{CNRS, Aix-Marseille Universit\'{e}, IM2NP (UMR 7334), Institut Mat\'{e}riaux Micro\'{e}lectronique et Nanosciences de Provence, Marseille, France.}

\author{I.~Chiorescu}
\email{ic@magnet.fsu.edu}
\affiliation{Department of Physics and The National High Magnetic Field Laboratory, Florida State University, Tallahassee, Florida 32310, USA}

\date{\today}%

\begin{abstract} The properties of rare-earth elements diluted in non-magnetic crystals make them a promising candidate for a quantum memory due to their limited Hilbert space. The control and readout of the qubit states require a highly sensitive measurement and large coupling of the spin ensemble with the electro-magnetic mode of a superconducting resonator. We report sensitive detection of forbidden transitions of electro-nuclear states from the minority species of $^{155}$Gd and $^{157}$Gd isotopes which opens the possibility of connecting quantum states with very different spin projections. Cavity perturbation measurements seen in the reflected signal allows the detection of about 7.6$\times10^{7}$ spins and the measurement of phase memory loss rate and spin-photon coupling strength.  \end{abstract}


\maketitle

\section{Introduction} \label{sec:introduction} 

The implementation of quantum information processors can benefit from the use of hybrid quantum systems that harness the strengths of multiple qubit platforms. On the one hand, superconducting qubits have fast logic-gate operations with high fidelity, while their performance is limited by the qubit decoherence times~\cite{Kjaergaard2020}. To circumvent this problem, hybrid systems combine superconducting circuits with other two-level systems (TLS) acting as a memory to store and retrieve the quantum information. Among the TLS, spin qubits are proposed as quantum memories given their long decoherence and relaxation times~\cite{Balasubramanian2009,Bertaina2020}. Such systems exploit the electron  or nuclear spin degree of freedom which can be coupled to resonant electromagnetic modes in superconducting or three-dimensional cavities~\cite{Blencowe2010}, acting as a bus between  the superconducting qubit and the spin ensemble\cite{Kubo2011}.

Many spin-ensemble-based systems have been investigated as prospects for storing quantum information including organic radical magnets \cite{Ghirri2016, Bonizzoni2017,Chiorescu2010a}, N-\textit{V} centers in diamond and other point defects in Si \cite{Kubo2010a,Schuster2010a,Amsuss2011,Bienfait2016,OSullivan2020}, and quantum spins in crystals~\cite{Bushev2011,Probst2013,Tkalcec2014,Jenkins2017,Bertaina2017, Bertaina2009, Orio2021}. Recent studies using Er$^{3+}$:Y$_2$SiO$_5$ spin-diluted crystals\cite{Probst2015prb}, N-\textit{V} centers in diamond\cite{Grezes2015} and Bi defects in Si \cite{Ranjan2020prl} demonstrated storing and retrieving the state of microwave photons at high power and near the quantum limit regime. 

Incorporating rare-earth (RE) ions in nonmagnetic matrices has generated considerable interest as potential candidates as well ~\cite{Bushev2011,Probst2013,Probst2014,Tkalcec2014,Wisby2014,Yue2017,Budoyo2018, Bertaina2007}. In particular, the \textit{S}-state lanthanides (Eu$^{2+}$, Gd$^{3+}$, Tb$^{4+}$) have zero total orbital angular momentum, which partially suppresses the spin-lattice and crystal-field interactions~\cite{Baibekov2017}. These large-spin ($S$ = 7/2) ions possess a rich energy-level structure that spans the Hilbert space of a three-qubit system allowing a quantum gate to be obtained by applying resonant $\pi$~pulses~\cite{leuenberger_quantum_2001,Jenkins2017}.

In the present report we study electronuclear transitions at the mesoscale level of the minority species of $^{155}$Gd$^{3+}$ and $^{157}$Gd$^{3+}$ (\textit{I}=3/2) isotopes,  in order to achieve local (on-chip) control of the electronic and nuclear degrees of freedom. In the Gd odd isotopes, the nonzero $I$ creates 2$I$+1 hyperfine split electronuclear states with transition frequencies close enough to be studied with the single mode of an on-chip resonator by means of field tuning of the resonance frequency\cite{Groll2010}. The Gd$^{3+}$ ions are diluted in a CaWO$_{4}$ single crystal and weakly coupled with the electromagnetic mode of a coplanar stripline superconducting resonator. The CaWO$_4$ host matrix provides a large crystal field that gives spin transitions at approximately 18~GHz. The microwave field distribution induced by the resonator geometry allows us to connect electronuclear quantum states with very different spin projections ($\Delta m_s\neq1$) that would otherwise be forbidden. The large values for the spin of a RE ion, cavity resonant frequency, and spin projection swing contribute to an increase of the overall spin-photon coupling strength. This makes the $^{155,157}$Gd$^{3+}$:CaWO$_4$ spin crystal an interesting candidate for a quantum hybrid system when integrated into current circuit quantum electrodynamics architectures.            

\section{Resonator design and experimental setup}

\begin{figure}[t]
	\centering
	\includegraphics[width=\mycol\columnwidth]{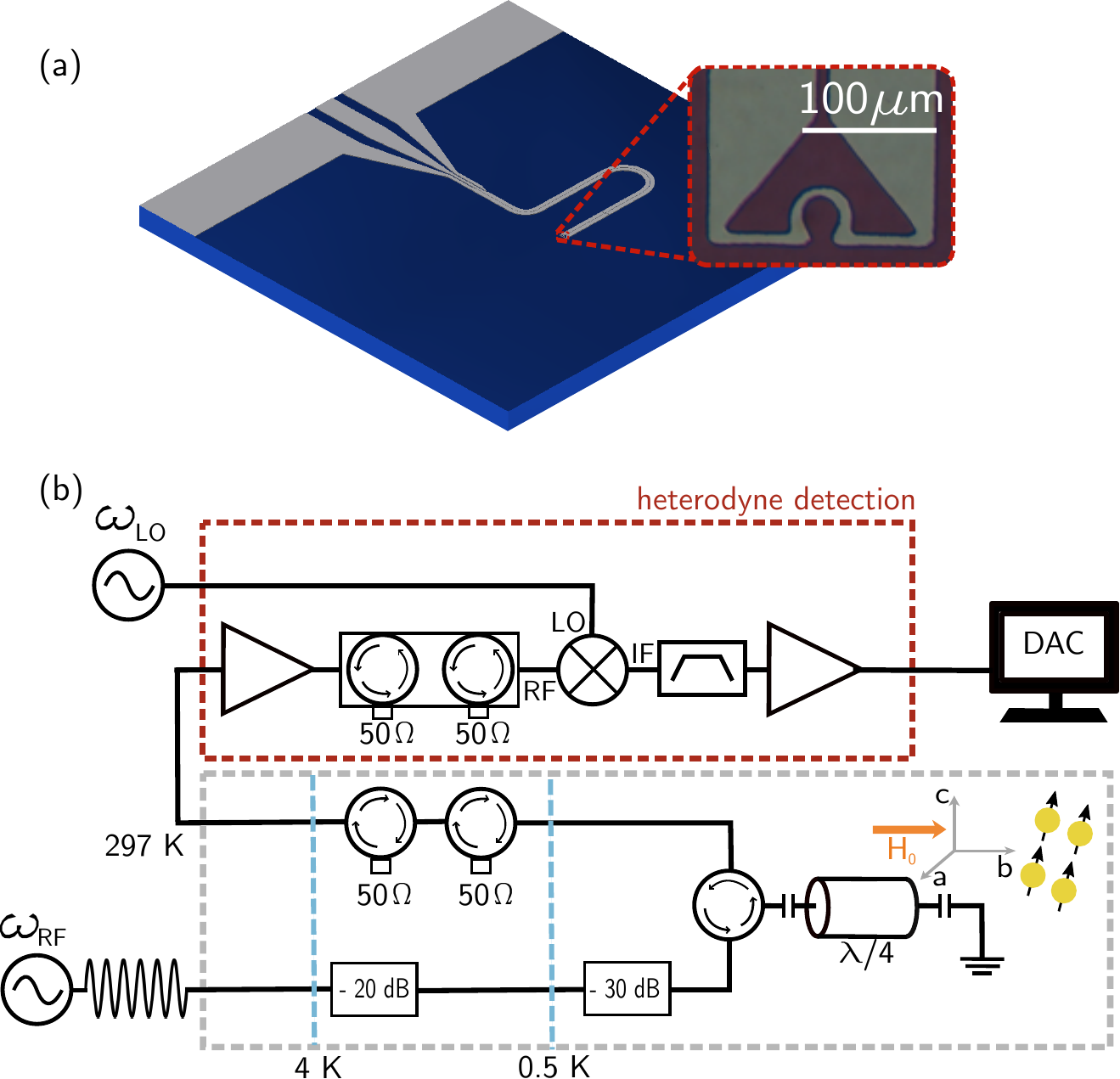}
	\caption{(a) Coplanar stripline resonator design; the inset shows a photograph of the $\Omega$-shaped termination. (b) Heterodyne detection schematics. An excitation pulse is generated by means of a MW synthesizer. The signal is attenuated at the different stages of a dilution refrigerator before reaching the on-chip resonator anchored at T~$\simeq$~0.38K. The reflected signal is then amplified at room temperature before down-mixing by a lower tone signal and recorded by the Digital-to-Analog Converter (DAC).} 
	\label{fig1}
\end{figure}

The on-chip microwave resonators are built from a 20-nm Nb thin film deposited in a Si/SiO$_2$ substrate (300~nm of SiO$_2$) using dc magnetron sputtering. For the resonator pattern standard UV lithography techniques and reactive ion etching in a CF$_4$-O$_2$ atmosphere are used to obtain the desired structure. The use of such thin films allows one to apply in-plane magnetic fields without considerably modifying the resonator quality factor $Q$, due to field-induced losses~\cite{Groll2010, Kwon2018, Kroll2019}. The resonator geometry has a designed transition from a coplanar waveguide to a coplanar stripline~\cite{Anagnostou2008} with all dimensional parameters corresponding to a 50-$\Omega$ impedance. A $\lambda/4$ resonator is realized by interrupting the stripline thus creating a capacitive coupling between the resonator and the transmission line. At the end of the resonator, a short-circuit termination is designed to have an $\Omega$~shape with internal radius of 15~$\mu$m (see Fig.~\ref{fig1}a). The microwave energy, and in particular its magnetic field component, is concentrated in this $\Omega$ loop and the Gd sample is to be placed directly on top.   

The resulting resonator and the sample holder are thermally anchored to the mixing chamber of a dilution refrigerator via a cold finger that centers the sample inside a large superconducting vector magnet. The temperature at the sample stage is kept at \textit{T}~$ \simeq $ 0.38~K. A schematic diagram of the full measurement setup is presented in Fig.~\ref{fig1}b. A long microwave (MW) pulse of 25~$\mu$s is sent by a synthesizer to probe the cavity-spin system. The signal is attenuated at different stages of the dilution refrigerator to reduce the thermal noise coming from room temperature. At the sample holder stage, the signal passes through a circulator allowing isolation of the reflected signal towards the output line. To excite different electronuclear transitions, a superconducting vector magnet is used allowing precise tuning of the in-plane field with a 0.2-G resolution. The reflected signal is amplified at room temperature by 35 dB in the 17$-$20 GHz range before being down-mixed with a lower frequency tone (LO). A band-pass filter centered at $\omega_{\text{RF}} - \omega_{\text{LO}}$ is used before further amplification. The signal is finally recorded and digitized by a high-speed acquisition card. 
\begin{figure}[t]
		\includegraphics[width=\columnwidth]{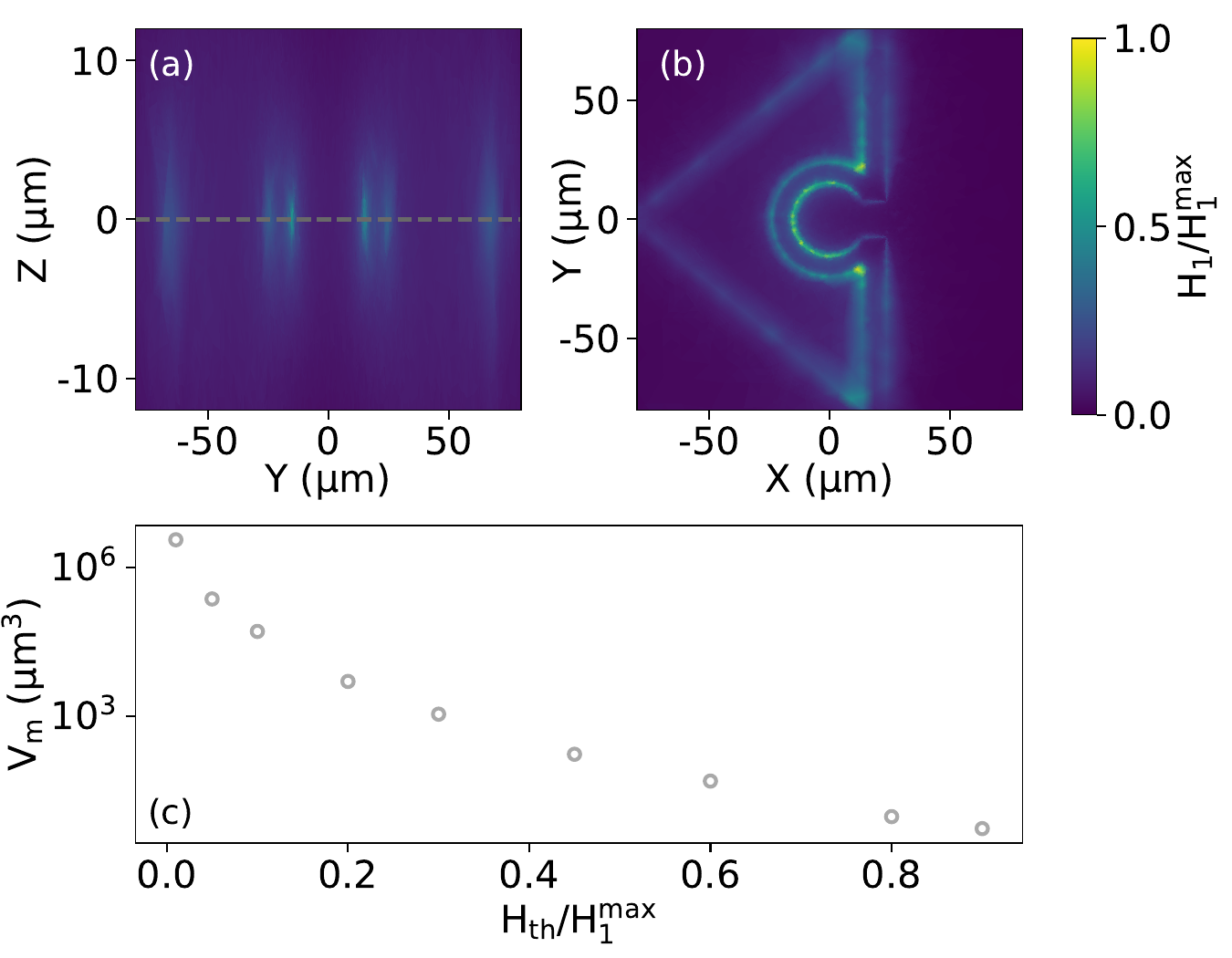}
		\caption{(a) FE simulation of the normalized $H_1$ field distribution at the center of the $\Omega$-shaped loop shows the field intensity drops by 70~\% above 3~$\mu$m of the chip surface ($Z$ = 0 $\mu$m). (b) Top view of the $\Omega$-shaped termination at the chip surface showing where the MW $H_1$ field is concentrated. (c) Volume calculation vs $H_{th}/H_{max}$; a mode volume of 1100 $\mu$m$^3$ is obtained in agreement with an estimation using eq. \ref{Vmode}.}
		\label{fig2}
	
\end{figure}

We can estimate the volume of the microwave magnetic field using COMSOL MULTIPHYSICS finite-element (FE) simulations. Evaluations of $H_1$ are shown in Fig.~\ref{fig2}a for a cut through the center of the $\Omega$-shaped termination and at the chip surface (Fig.~\ref{fig2}b). Using the full chip geometry, one can identify the volume $V_m$ containing $H_1$ field ranging from the maximum value $H_1^{max}$ near the surface to a certain threshold $H_{th}$ as shown in Fig.~\ref{fig2}c as function of $H_{th}/H_1^{max}$. Following the definition of a mode volume~\cite{AGRAWAL201327,ansyslumerical}, $(\int H_1^2 dV)^2/\int H_1^4 dV$, one can get an approximate value using the equation

\begin{equation}
V_m = \frac{\left(\int_{V_{sample}} H_1^2\,dV\right)^2}{\int_{V_{cavity}} H_1^4\,dV},
\label{Vmode}
\end{equation}
where $V_{sample}$ and $V_{cavity}$ are the sample and full chip volumes, respectively. One obtains a mode volume of about 1100 $\mu$m$^3$, similar to the value at $H_{th}/H_1^{max} = 0.3$. Further away from this threshold, the magnetic energy density drops to approximately 5\% of that near the $\Omega$~loop.

\section{Field dependent spectroscopic measurements} \label{sec:field-dependent-spectroscopic-measurements}

\begin{figure}[t]
	\centering
	\includegraphics[width=\mycol\columnwidth]{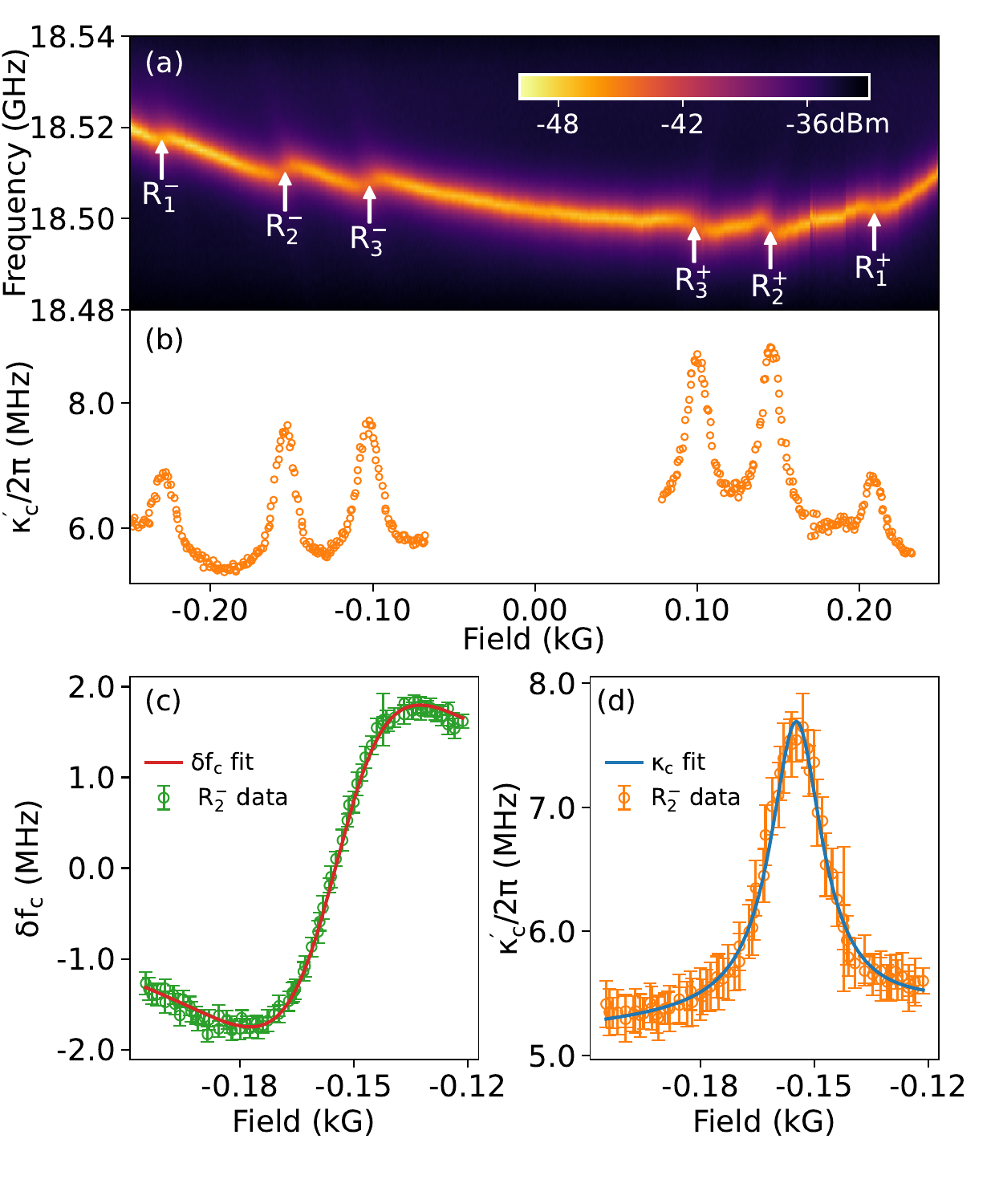}
	\caption{(a) Field-dependent cavity spectroscopy signal. Interactions between the spins and the cavity mode are highlighted by the white arrows. (b) Measured field-dependent cavity width, $\kappa_c^{\prime}$, across each resonance. The cavity width is obtained from fitting the reflection amplitude parameter $S_{11}$ to the frequency sweep cross section of the contour scan (see ref. \cite{Franco-RiveraSI}). (c) Measured cavity resonance frequency perturbation $\delta f_c$ and (d) cavity widths $\kappa_c^{\prime}$ for the spin transition labeled $R_2^{-}$ represented by the navy and red circles respectively. Fits to the measured resonance frequency and cavity width perturbation are performed using Eqs.~(\ref{omegaceq}) and (\ref{kappaceq}).} 
	\label{fig3}
\end{figure}

A single crystal of Gd$^{3+}$:CaWO$_4$ with a 0.05\% spin concentration is placed on top of the $\Omega$~loop of the Nb resonator and the setup described above is used to record the reflected signal as a function of field and frequency. At resonance, the microwave frequency $\omega$ matches the spin-level splitting and the absorbed power generates a dip in the reflected signal. The electronuclear levels are eigenvalues of the Gd$^{3+}$ $S=7/2$ ion in the $I4/a$ tetragonal symmetry\cite{Zalkin1964} of the host matrix \cite{Hempstead1960}:

\begin{equation}
\begin{split}
\mathcal{H} =& \mu_B \boldsymbol{H}^{T}_0 \boldsymbol{g}\boldsymbol{S} + \boldsymbol{S}^{T}\boldsymbol{A}\boldsymbol{I}\\
&+ B_2^0 O_2^0 + B_4^0 O_4^0 + B_4^4 O_4^4 + B_6^0 O_6^0 + B_6^4 O_6^4 
\end{split}
\label{hamiltonian}
\end{equation}

where $O_p^k$ are the Stevens operators \cite{Rudowicz2004}, $\mu_B$ is the Bohr magneton, $\mathbf{H}_0$ is the applied magnetic field vector and ($g_{\|}$, $g_{\perp}$) = (1.991,1.992) are the $g$-factors parallel and perpendicular to the crystallographic \textit{c}~axis respectively. The presence of the \textsuperscript{155}Gd ($A_{155}$ = 12.40~MHz abundance 14.7\%) and \textsuperscript{157}Gd ($A_{157}$ = 16.28~MHz abundance 15.7\%) \cite{Hempstead1960,Baibekov2017} isotopes with nuclear spin \textit{I} = 3/2 causes the splitting of each electronic \textsuperscript{8}\textit{S}\textsubscript{7/2} multiplet into $2I+1$ states. As discussed below, our spectroscopically found values for the prefactors of the Stevens operators are within an approximately 2~\% difference of values obtained in previous studies \cite{Harvey1971}. $\mathbf{H}_0$ provides an \textit{in-situ} tool to control the admixture of states of different $S_z$ projections. In the present study, the field is applied almost perpendicular to the $c$~axis, as discussed below, leading to a significant mixture of pure $|S_z\rangle$ states. Some of the eigenstates are shown in the Supplemental Material \cite{Franco-RiveraSI}. 

Measurements of the reflected microwave power as a function of frequency and magnetic field are given in Fig.~\ref{fig3}a; for a fixed field, the excitation signal is swept around the cavity resonance and the process is repeated from negative to positive values of the magnetic field. The contour plot identifies six spin resonances visible as a perturbation of the cavity resonance frequency. As a side note, one observes that the field dependence of the cavity resonance is not a quadratic decrease with field~\cite{Yip1992,Groll2010}. This could be due to how the main nuclear Gd$^{3+}$ species, $I=0$, couples with the cavity; classically, the sample magnetization in field can change the resonator inductance. Nevertheless, this aspect is of secondary importance to the current study which focuses on the electronuclear transitions of the minority species $I=3/2$.

The perturbation of the cavity properties by a spin transition is significant in the case of superconducting planar cavities, due to a reduced mode volume, leading to a large filling factor. In the limit of weak spin-photon coupling, the effect of dissipation can be represented by using a complex frequency $\overline{\omega}_c = \omega_c - i\left(\omega_c/2Q\right) = \omega_c - i\kappa_c$. By introducing a magnetic sample in the cavity, the resonator inductance changes and the complex resonance frequency becomes $\delta \overline{\omega}_c = \delta \omega_c - i\delta \kappa_c$. The relative shift in frequency is proportional to the sample's susceptibility $\delta \overline{\omega}_c/\omega_c = -\beta \overline{\chi} = -\beta\left(\chi' -i\chi'' \right)$, $\beta$ being a geometric factor that depends on the spin distribution around the resonator volume and its coupling to the cavity mode~\cite{Donovan1993,Bushev2011}. By analyzing the real and imaginary part of the expression, one can find the dependence on frequency. It is important to note that even in the weak coupling regime, large filling and $\beta$ factors can lead to perturbations so large that the $Q$ factor becomes impossible to measure.  
\begin{figure}[t]
	\centering
	\includegraphics[width=\mycol\columnwidth]{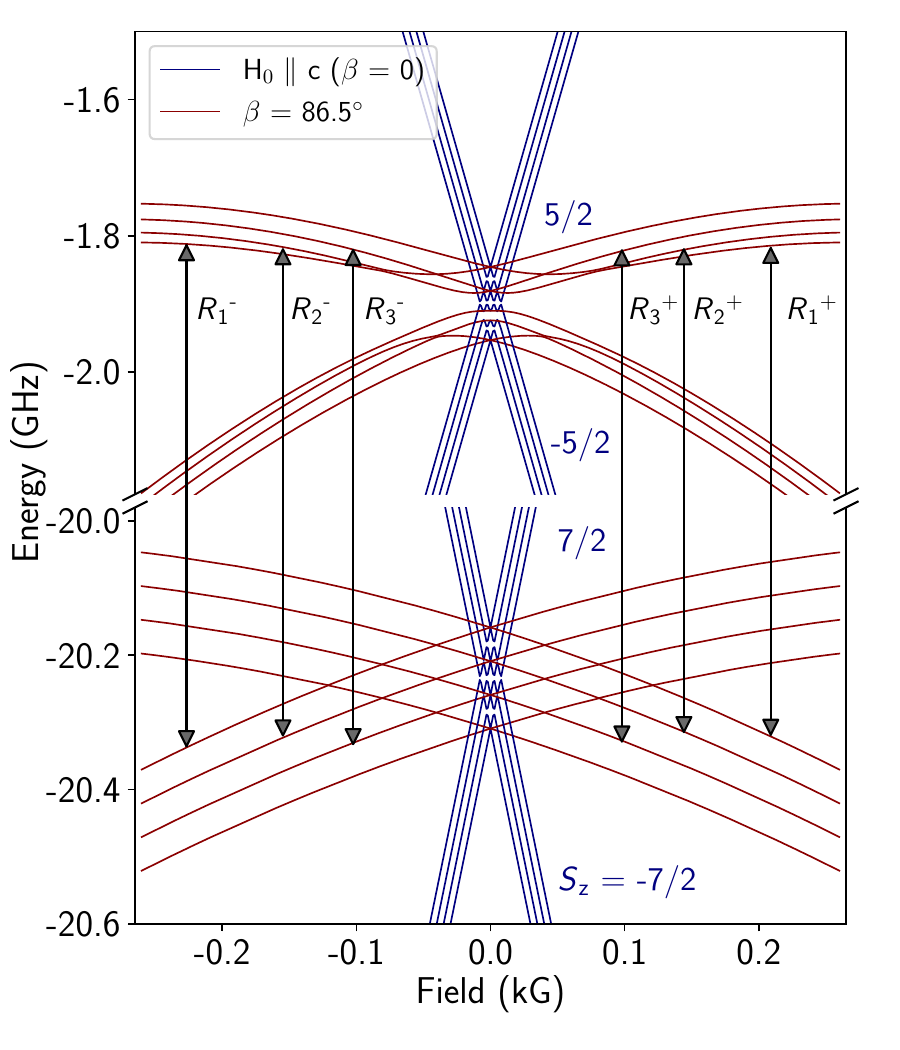}
	\caption{Energy levels of the Gd$^{3+}$ spin Hamiltonian obtained from diagonalization for fields applied parallel (navy, $\beta$~=~0\textdegree) and close-to-perpendicular (dark red, $\beta$~=~86.5\textdegree) to the crystallographic \textit{c}~axis. The observed transitions are shown by arrows and arise from the hyperfine splitting ($\Delta I_z = 0$) of the ground and third excited state of the $^{7/2}$\textit{S}$_8$ electronic multiplet.} 
	\label{fig4}
\end{figure}

In a similar way, the following expressions derived from the input-output formalism in the presence of a spin system~\cite{herskind2009} describe the perturbation of the cavity resonance and width:
\begin{gather}
\omega_c' = \omega_c + \frac{g_c^2\Delta}{\Delta^2 + \gamma_s^2}, \label{omegaceq} \\
\kappa_c'=\kappa_c + \frac{g_c^2 \gamma_s}{\Delta^2 + \gamma_s^2}, 
\label{kappaceq}
\end{gather}
where $\omega_c$ and $\kappa_c$ are the unperturbed cavity resonance frequency and width, respectively, $\gamma_s$ is the spin's phase memory loss rate, $g_c$ is the cavity spin-ensemble coupling constant and detuning $\Delta=\omega-\omega_s$ is defined as the difference between the probing frequency $\omega$ and the level separation $\omega_s$ between the two electronuclear levels under investigation. Consequently, the power reflected by the cavity can be written as (see also~\cite{Abe2011,Angerer2016}) 
\begin{equation}
	\left|S_{11}\right|^2= \left|1+\frac{\kappa_e}{i(\omega-\omega_c')-\kappa_c'}\right|^2
	\label{s11eq}
\end{equation}
where $\kappa_e$ is the dissipation due to external factors and in practice is approximately equal to $\kappa_c$ or $\kappa_c'$. Eq.~\ref{s11eq} can reproduce a double-dip $\omega$ scan characteristic of the strong-coupling regime for a cooperative factor $C = g_c^2/\kappa_c \gamma_s \gtrsim 1$. For $C\ll1$, $\left|S_{11}(\omega)\right|^2$ has a Lorentzian behavior with a perturbed center and width, as in this study.
 
The reflected power measurements presented in Fig.~\ref{fig3}a are fitted with Eq.~\ref{s11eq} for each field. Details of the fitting procedure and a representative example are provided in the Supplemental Material ~\cite{Franco-RiveraSI}. The process is repeated for each field and the obtained dependence of $\frac{\kappa_c'}{2\pi}$ on field is shown in Fig.~\ref{fig3}b. Similarly, the field dependence of the perturbed $\frac{\omega_c'}{2\pi}$ is obtained. To extract the values of the $\frac{g_c}{2\pi}$ and $\frac{\gamma_s}{2\pi}$ parameters, fits using Eqs.~\ref{omegaceq} and \ref{kappaceq} as a function of detuning $\Delta$ are needed. As detailed in the Supplemental Material~\cite{Franco-RiveraSI}, numerical diagonalization of the spin Hamiltonian is required to established the relationship between $\omega_s$ and the applied field. This is particularly of essence in the case studied here, where the static field is applied perpendicularly to the $c$~axis to give a finite probability to transitions otherwise considered forbidden. Note that while $\Delta$ is controlled by magnetic field, the perturbation analysis provides $\frac{\gamma_s}{2\pi}$ and $\frac{g_c}{2\pi}$ in hertz. Examples of fits using Eqs.~(\ref{omegaceq}) and (\ref{kappaceq}) (to which a linear background is added; see Suplemental Material for details) are shown in Figs.~\ref{fig3}(c) and \ref{fig3}(d).      
\begin{table}
	\caption{Spin ensemble decoherence rate $\gamma_s$ and spin-photon effective coupling constant $g_c$ for each of the six transitions (labels as in Fig.~\ref{fig3}a) from perturbed cavity width and resonance frequency fits. The nuclear quantum number $I_z$ of levels involved in the transition is given as well.}
	\begin{tabular}{cccccc}
		\hline\hline
		& \hspace{3mm}$I_z$\hspace{3mm} & \hspace{4mm}$\gamma_s/2\pi$\hspace{4mm} & \hspace{4mm}$g_{c}/2\pi$\hspace{4mm} & \hspace{4mm}$\gamma_s/2\pi$\hspace{4mm}& \hspace{4mm}$g_{c}/2\pi$\hspace{4mm}\\
		&& \multicolumn{2}{c}{(MHz)}& \multicolumn{2}{c}{\hspace{4mm}(MHz)}\\
		& & \multicolumn{2}{c}{cavity width} & \multicolumn{2}{c}{freq. shift}\\
		\hline
		$R_1^{-}$ & -3/2& 7.3$\pm$0.4 & 3.0$\pm$0.1 & 11.0$\pm$0.6 & 3.3$\pm$0.1\\
		$R_1^{+}$ &+3/2 & 7.6$\pm$0.5 & 2.9$\pm$0.1 & 9.4$\pm$2.2 & 2.7$\pm$0.2\\
		$R_2^{-}$ &-1/2 & 8.1$\pm$0.25 & 4.4$\pm$0.1 & 19.9$\pm$0.25 & 8.0$\pm$0.1\\
		$R_2^{+}$ &+1/2 & 9.0$\pm$0.4 & 5.0$\pm$0.1& 10.2$\pm$0.5 & 5.2$\pm$0.1\\
		$R_3^{-}$ & +1/2& 9.8$\pm$0.3 & 4.7$\pm$0.1 & 18.3$\pm$0.4 & 6.4$\pm$0.1\\
		$R_3^{+}$ & -1/2 & 10.0$\pm$0.4 & 5.1$\pm$0.1 & 15.5$\pm$1.2 & 5.6$\pm$0.2\\
		\hline  
	\end{tabular}
	\label{table1}
\end{table}

The field dependence of $\omega_s$ for each studied resonance is found using the EasySpin package~\cite{Stoll2006}. A single value of \textit{A} = 14.34~MHz is used for simplicity since the field separation between odd-isotope resonances lies well within the resolution of our spectroscopic measurements. The crystal field parameters $B_2^0$= -938.4, $B_4^0$=-1.247, $B_4^4$=-7.305, $B_6^0$ = 5.712~$\times$10$^{-4}$, and $B_6^4$ = 70.0~$\times$10$^{-4}$ (all in MHz) are determined by least-squares comparison between the experimental resonance fields and fields obtained from numerical diagonalization of the spin Hamiltonian.

The obtained eigenvalues of $\mathcal{H}$ are presented in Fig.~\ref{fig4} for applied fields parallel (navy) and close-to-perpendicular (dark red) to the \textit{c}~axis. For parallel $\mathbf{H}_0$-fields the calculated eigenvalues are characterized by good quantum numbers $S_z$ and split by the crystal field for $\mathbf{H}_0=0$. A static field almost perpendicular to the easy axis ($\beta\sim 90^{\circ}$ and fine tuned in EasySpin) causes an admixture of the $S_z$ states between the electronic multiplet ground state and its third excited state, causing a swing in spin projection of approximately $4-5$ while $\Delta I_z = 0$ (see Supplemental Material~\cite{Franco-RiveraSI} for tablulated values). The aforementioned transitions are highly forbidden in the context of conventional ESR where $\Delta S_z = 1$. The observed resonances are highlighted in Fig.\ref{fig4} with gray arrows. Only the three highest-probability transitions of the $2I+1$ hyperfine resonances are observed as can be seen in the field-dependent spectroscopic measurements of Fig.~\ref{fig3}.
    
\section{Decoherence rates and spin sensitivity} 
\label{sec:decoherence-rates-and-spin-sensitivity}
    
\begin{figure}
		\includegraphics[width=\columnwidth]{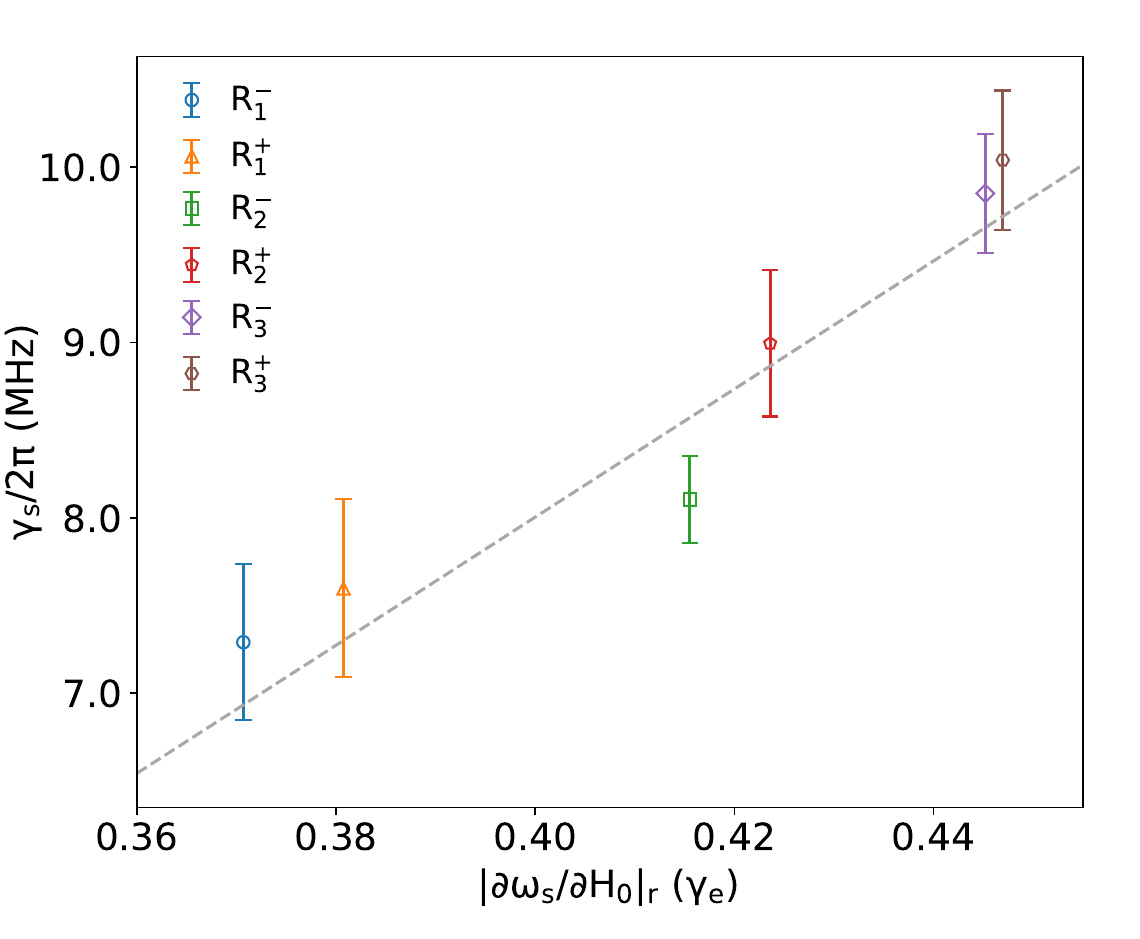}
	\caption{Spin ensemble decoherence rate $\gamma_s/2\pi$ from cavity width fits versus $\left|\partial \omega_s / \partial H_0 \right|_{r}$.}
	\label{fig5}
	
\end{figure}

From the fits of the six resonances, similar to those shown in Fig.~\ref{fig3}(c) and \ref{fig3}(d), the $\gamma_s$ rate and the cavity-spin-ensemble coupling strength $g_c$ are obtained. The values are listed in Table~\ref{table1} in increasing order of the field derivative $\left|\partial \omega_s / \partial H_0 \right|_{r}$ of $\omega_s$ at resonance $\Delta=0$. In the case of the linewidth fit (Eq.~\ref{kappaceq}), its peak is located over a narrower region of the field and therefore the fit reliability is less sensitive to background variations. In contrast, the $\omega'_c$ function covers a large field span around a resonance and can be affected by the nonlinear background of $\omega_c(H_0)$. The values of $\gamma_s$ obtained from the $\kappa_c'$ fit are shown in Fig.~\ref{fig5} as a function of the field derivative, where $\left|\partial \omega_s / \partial H_0 \right|_{r}$ is given in units of $\frac{\gamma_e}{2\pi}=2.8025$~MHz/G, the gyromagnetic ratio of the free electron. The positive correlation between decoherence rate and slope agrees with the known view that decoherence is improved when field noise is canceled in first order (so-called clock transitions). The rates $\gamma_s$ obtained from the resonance frequency perturbation are fluctuating around a weighted average value of $12.2\pm1.6$~MHz but are likely to be more affected by systematic errors as mentioned above. Other fit parameters are the unperturbed cavity resonance $\omega_c$, which follows the nonlinear background (see Fig.~\ref{fig3}), and the unperturbed cavity width $\kappa_c$. The latter gives an average value $\frac{\kappa_c}{2\pi}=5.84\pm0.13$~MHz leading to a cooperativity factor \textit{C} with an overall range from 0.15 to about 0.45.

The value of decoherence rate $\gamma_s$ compares well against continuous wave and transient pulse spin echo measurements performed at 6~K~\cite{Baibekov2017}, which show similar spectral linewidths  14$-$42~MHz ($T_2^{\star}$ = 25$-$70~ns) while the phase memory times are significantly longer ($T_2\approx5-10\mu$s). The nature of the linewidth is inhomogeneous~\cite{Baibekov2017} but the long spin-echo time is instrumental in reducing the decoherence effects and the spin sensitivity per $\sqrt{Hz}$. The phase memory loss rate can be attributed to intrinsic mechanisms such as spin-lattice relaxation and dipolar interactions with neighboring Gd or $^{183}$W isotopes and, most importantly, to random distribution of electric field effects on the crystal-field spin Hamiltonian \cite{Baibekov2011}. The Gd$^{3+}$ replaces Ca$^{2+}$ causing a random distribution of internal charge compensation regions. To fully elucidate the nature of the intrinsic decoherence mechanisms at sub-kelvin temperatures, spin-echo or Carr-Purcell-Meiboom-Gill pulse sequences are to be explored.

For the given Gd-doped concentration and effective cavity volume a rough estimation of the total number of spins is $N_{tot}=1.78\times10^{9}$ and the number of spins excited per transition is obtained as in Ref. \cite{Bienfait2016}, $N_s=N_{tot}\left[0.304/(2I+1)\right]\kappa_c/\gamma_s=7.55 \times 10^{7}$ spins, where 30.4\% is the natural abundance of Gd odd isotopes. In order to estimate the number of spins that contribute to the signal, the spin-photon coupling strengths are calculated for the different mode volumes shown in Fig. \ref{fig2}c as detailed in Supplemental Material Sec. II~\cite{Franco-RiveraSI}. The numerical analysis, based on COMSOL and EasySpin simulations, reveals that the mode volume is actually well approximated by Eq.~\ref{Vmode} and the most important microwave field component is the one $\perp c$~axis. It is worth noticing that $N_s$ is of the same order of magnitude as the one obtained with sensitive spin detection techniques based on direct current - Superconducting QUantum Interference Device (dc-SQUID) loops~\cite{Yue2017, Budoyo2018, Toida2016} and other techniques~\cite{Eichler2017,Weichselbaumer2019}. Using an average value of $g_c/2\pi=4.2 \pm 0.4$~MHz from cavity width measurements, the average single spin coupling can be estimated as $g_0/2\pi=g_c/2\pi/\sqrt{N_s}\approx 480 \pm 50$ Hz. In our study, $N_s$ represents the total number of spins obtained in a continous-wave experiment (large bandwidth). The spin sensitivity per $\sqrt{Hz}$ can be improved by the use of quantum limited parametric amplification combined with the use of advanced pulsed sequences \cite{Bienfait2016,Eichler2017} and including submicrometer-scale constrictions at the anti-node location as done in the case of \textit{LC} planar microresonators with nanometer-wide inductance strip~\cite{Probst2017, Ranjan2020}. 

CaWO$_4$ is an excellent host for a spin-diluted quantum memory due to its low nuclear magnetic moment density (only $^{183}$W carries $I=1/2$ with natural abundance 14\%) and thus low spin bath fluctuations. Recently~\cite{leDantec2021}, Er$^{3+}$ ($S=1/2$) ions diluted in CaWO$_4$ demonstrated a 23-ms coherence time measured by a single Hahn echo sequence at millikelvin temperatures using an \textit{LC} resonator patterned directly on the crystal. It is likely that the Gd$^{3+}$ ions have larger resonator-spin coupling due to a larger magnetic moment and, at the same time, a larger operating Hilbert space since $S=7/2$.

\section{Conclusion} 
\label{sec:conclusion}

To conclude, we present the coupling of the electromagnetic mode of a superconducting $\Omega$~loop with the electronnuclear states of $^{155,157}$Gd isotopes diluted in a nonmagnetic crystal. Highly sensitive measurements of cavity perturbation in reflected signal make possible the study of forbidden transitions in transverse field. Spin phase memory loss rates and spin-photon coupling constant are obtained for six hyperfine resonance lines.  These results present an opportunity for on-chip control of an electronnuclear system with close transition frequencies and limited Hilbert space for quantum operations.      

\section*{Acknowledgments}
This work was performed at NHMFL at the Florida State University and supported by the National Science Foundation through Grant No. NSF/DMR-1644779 and the State of Florida. We acknowledge discussions with Dr. Petru Andrei (FSU) and we thank Dr. A.M. Tkachuk for providing the sample. S.B. acknowledges support from the CNRS research infrastructure RENARD (award number IR-RPE CNRS 3443). L.C. acknowledges partial support by the Frontier Science Key Programs and the Young Investigator Program of the CAS (Grants No. QYZDY-SSW-JSC033 and No. 2016217) and the National Natural Science Foundation of China (Grants No. 62071458 and No. 11827805).

\bibliography{cpsresref.bib}

\end{document}


\title{Supplementar Material: On-Chip Detection of Electronuclear Transitions in the $^{155,157}$Gd Multilevel Spin System}

\author{G.~Franco-Rivera}
\email{gfranco@magnet.fsu.edu}
\affiliation{Department of Physics and The National High Magnetic Field Laboratory, Florida State University, Tallahassee, Florida 32310, USA}

\author{J.~Cochran}
\affiliation{Department of Physics and The National High Magnetic Field Laboratory, Florida State University, Tallahassee, Florida 32310, USA}

\author{L.~Chen}
\affiliation{Center for Excellence in Superconducting Electronics, Shanghai Institute of Microsystems and Information Technology, Chinese Academy of Science, Shanghai, China.}

\author{S.~Bertaina}
\affiliation{CNRS, Aix-Marseille Universit\'{e}, IM2NP (UMR 7334), Institut Mat\'{e}riaux Micro\'{e}lectronique et Nanosciences de Provence, Marseille, France.}

\author{I.~Chiorescu}
\email{ic@magnet.fsu.edu}
\affiliation{Department of Physics and The National High Magnetic Field Laboratory, Florida State University, Tallahassee, Florida 32310, USA}

\maketitle

\section{Cavity field dependent spectroscopy fitting procedure.} \label{sec1}

The field and frequency dependence of the cavity reflection signal is characterized by measuring the $|S_{11}|^2$ signal. First, we note that in the presence of spins, the cavity resonance frequency and width are perturbed and given by, respectively~\cite{herskind2009}:
\begin{gather}
\omega_c' = \omega_c + \frac{g_c^2\Delta}{\Delta^2 + \gamma_s^2}, \label{omegaceq} \\
\kappa_c'=\kappa_c + \frac{g_c^2 \gamma_s}{\Delta^2 + \gamma_s^2}, 
\label{kappaceq}
\end{gather}
where $\Delta=\omega-\omega_s$ is the detuning from the spin resonance, $\omega$ is the probing frequency, $\omega_s$ is the spin energy separation, $g_c$ is the overall spin-photon coupling parameter, $\gamma_s$ is the spin-ensemble decoherence rate $\left(1/T_2^{\star}\right)$ and $\kappa_c, \kappa_c',\omega_c,\omega_c'$ are the unperturbed/perturbed cavity width and resonance frequency respectively.

Therefore, the expression derived from the input-output formalism in the presence of a spin system for a perturbed $|S_{11}|^2$ becomes~\cite{Abe2011,Angerer2016}:
\begin{equation}
	\left|S_{11}\right|^2  = \left|1+\frac{\kappa_e}{i(\omega-\omega_c')-\kappa_c'}\right|^2,
	\label{S11}
\end{equation}
which can be rewritten as $|S_{11}|^2=1+\kappa_e\frac{\kappa_e-2\kappa_c'}{(\omega-\omega_{c}')^2+\kappa_c'^2}$ where $\kappa_e$ is the dissipation due to external factors and in practice is approximately equal to $\kappa_c$ or $\kappa_c'$.
Experimentally, the parameters $\omega$ and $\omega_s$ are varied and a 2D representation of $|S_{11}(\omega,\omega_s)|^2$ can be obtained. To obtain the experimental parameters $g_c$ and $\gamma_s$, a two-step process is applied as decribed below. First, for each value of $\omega_s$ a fit based on Eq.~\ref{S11} is performed, to obtain the $\omega_s$ dependence of $\omega_c'$ and $\kappa_c'$. Second, for a given resonance the probe frequency $\omega$ is considered constant and equal to $\omega_c$. Eqs.~\ref{kappaceq} are used to fit $\omega_c'$ and $\kappa_c'$. The later step produces as output the values of $g_c$ and $\gamma_s$.
\begin{figure}[t]
	\centering
	\includegraphics[width=\mycol\columnwidth]{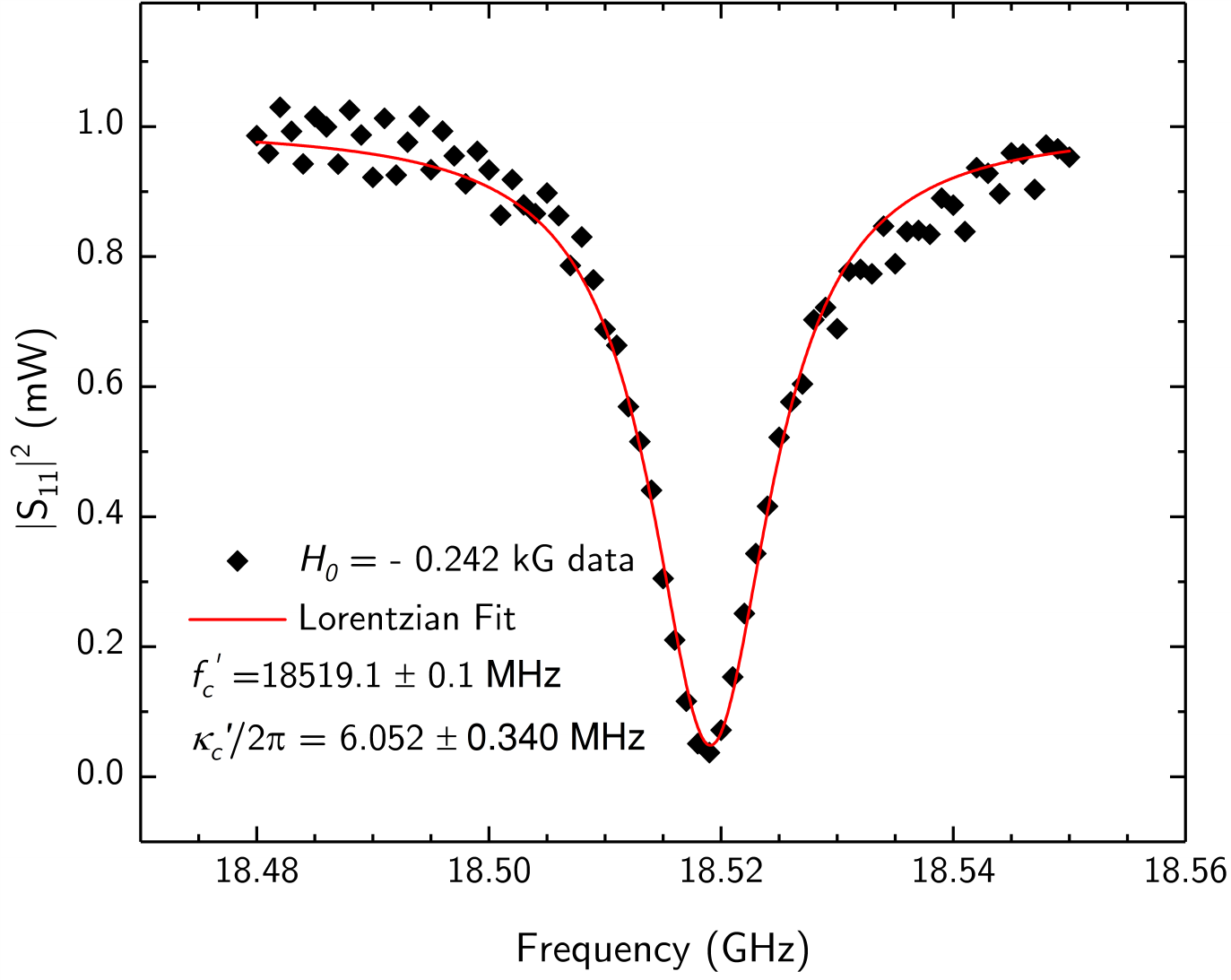}
	\caption{Resonator $|S_{11}|^2$ (black diamonds) as a function probe frequency. The Lorentzian fit (in red) gives $\omega_c'/2\pi$ and $\kappa_c'/2\pi$.} 
	\label{figS1}
\end{figure}
 
\begin{figure}[t]
	\centering
	\includegraphics[width=\mycol\columnwidth]{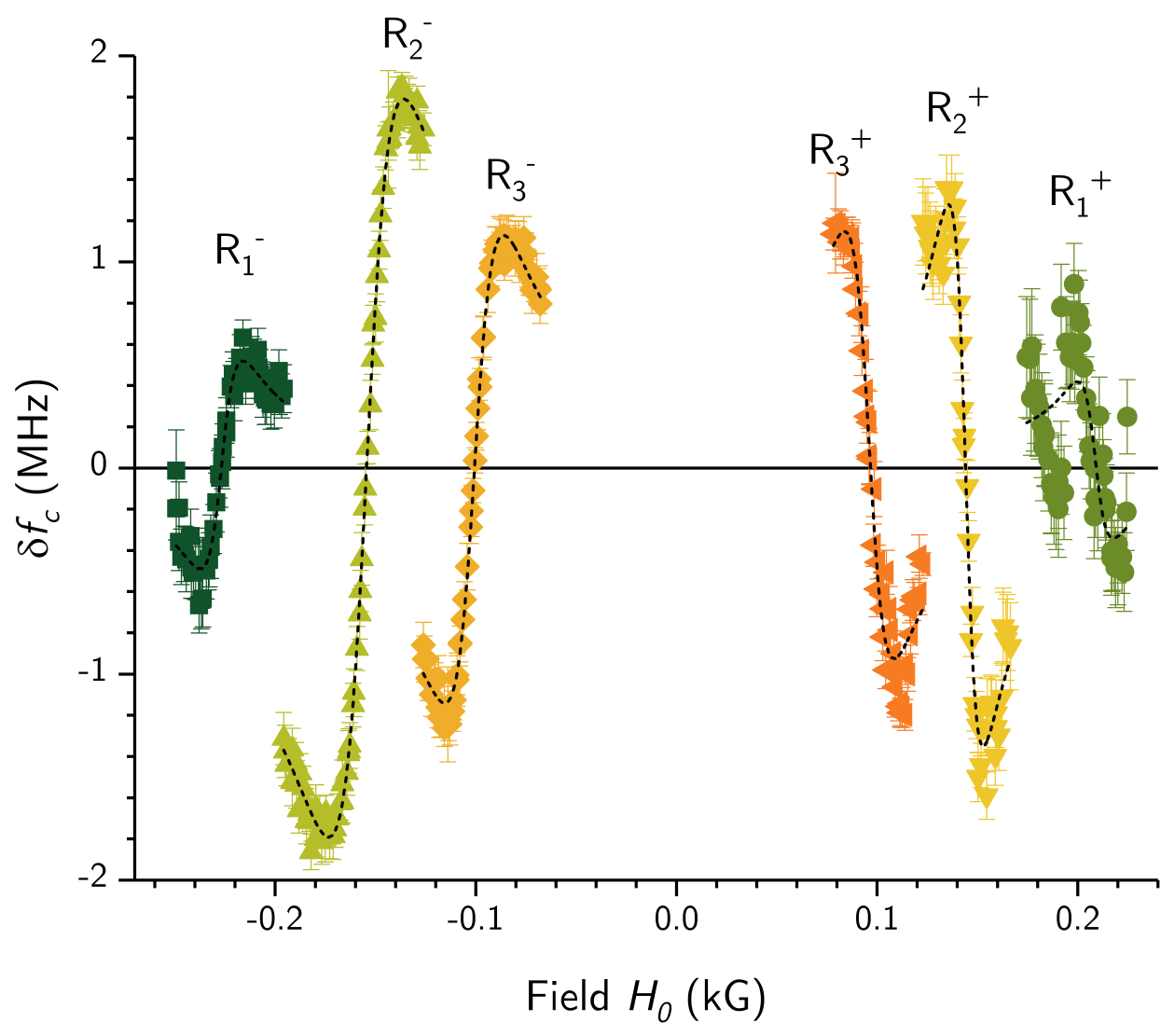}
	\caption{Resonator frequency shift, $\delta f_c$ as a function of the applied static field $H_0$ for the six observed resonances. Fits to the frequency shift data using Eq.~\ref{omegacfit} are respresented by the black dashed lines.} 
	\label{figS2}
\end{figure}

It is worth noting that Eq.~\ref{S11} reproduces a double-dip profile for $|S_{11}|^2$ if the strong coupling regime is attained. Otherwise, the profile is a Lorentzian with perturbed center and linewidth (the case of our study). The measured $|S_{11}|^2$ reflected signal (in mW units) was fitted with a Lorentzian lineshape as shown in Fig.~\ref{figS1} of the form:
\begin{equation}
	|S_{11}(f)|^2 = c - \frac{A}{2\pi} \frac{\frac{\kappa_c'}{2\pi}}{\left(f - f_c'\right)^2 + \left(\frac{\kappa_c'}{2\pi}\right)^2},
\end{equation}
where for fit simplicity and feasibility we approximate $\kappa_e-2\kappa_c'\approx-\kappa_c'$ in the denominator, $c$ is the baseline, $f_c' = \omega_c'/2\pi$, $A/2\pi$ and $\kappa_c'/2\pi$ are fit parameters in Hz ($\pi A$ represents the area of the Lorentzian while $\kappa_c'/2\pi$ is its half-width at half-max, HWHM). An example of such fit (in red) is given in Fig.~\ref{figS1} showing measured reflected power as a function of probing frequency. Similar fits are performed for all magnetic fields and parameters $f_c'$ and $\kappa_c'$ are obtained as a function of field. Since the applied field gives the energy separation between two levels at resonance $\omega_s$, this procedure is equivalent to finding $f_c'(\omega_s)$ and $\kappa_c'(\omega_s)$.

\begin{figure*}[t]
	\centering
	\includegraphics[width=\textwidth]{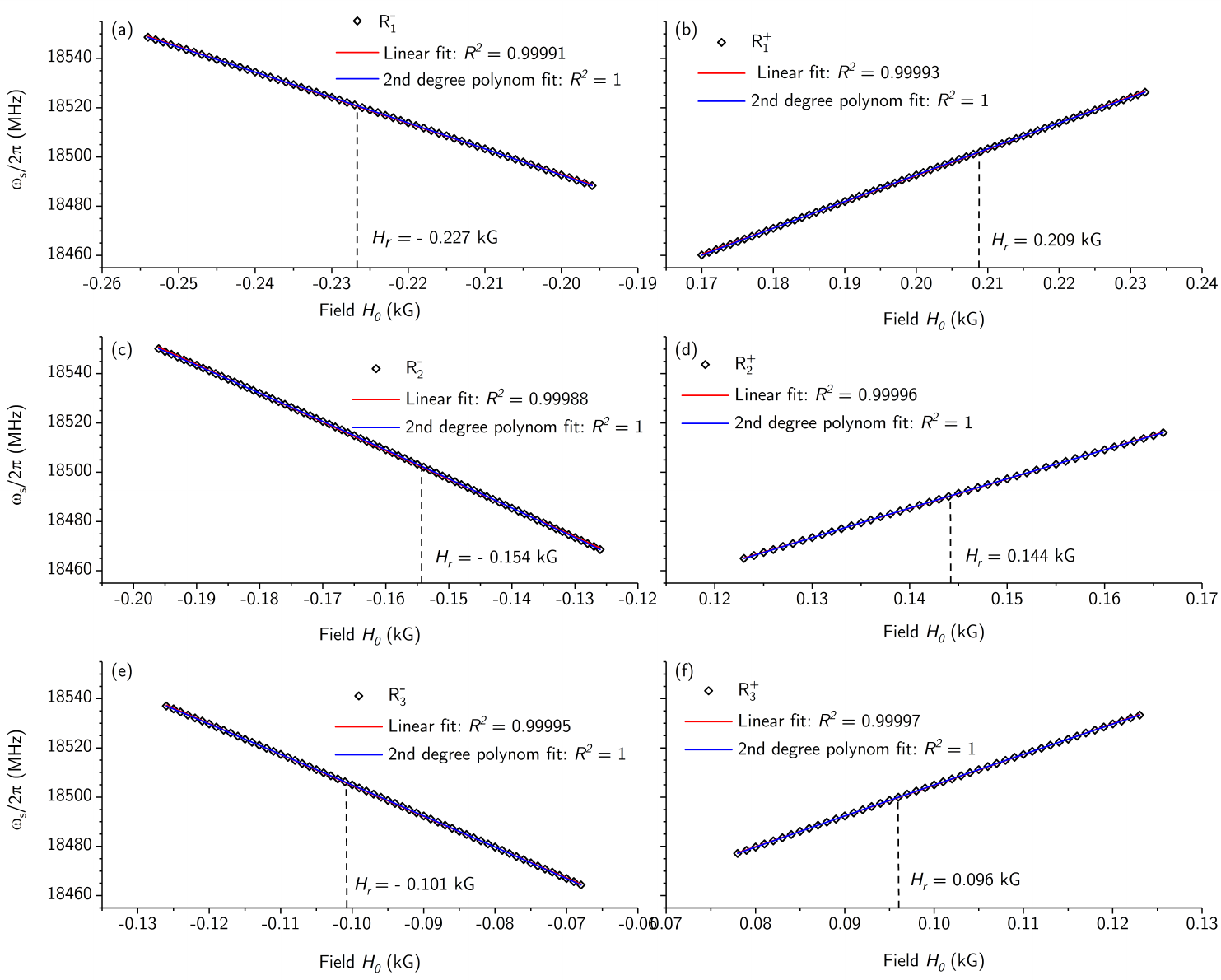}
	\caption{Field dependence of the spin transition frequencies $\omega_s/2\pi$ for all observed resonances represented by the black diamonds. The solid red line represent modeling the field dependence with a linear formula whereas the blue line represents a quadratic field dependence model leading to R$^2$=1.} 
	\label{figS3}
\end{figure*}

The spin-ensemble decoherence rate and effective coupling constant for each transition are then obtained by fitting $f_c'$ and $\kappa_c'$ as a function of field around a given resonance. In order to fit the data with Eqs. \ref{omegaceq} and \ref{kappaceq} a precise field dependence of the spin resonance detuning $\Delta$ is needed. For a free electron, $\Delta/2\pi=\gamma_e/2\pi \left(H_r - H_0\right)$ where $\gamma_e/2\pi=2.8025$~MHz/G is the gyromagnetic ratio of the free electron, $H_r$ is the resonance field and $H_0$ is the applied field. In general, a first order approximation of the field dependence is $\Delta = \omega_{c}(H_r)-\omega_s(H_0)=\partial\omega_s/\partial H_0\left(H_r-H_0\right)$ where $\partial\omega_s/\partial H_0$ can be obtained numerically using exact diagonalization of spin Hamiltonian. In the present study, we applied static fields almost perpendicular to the crystal c-axis causing admixture of states with different $|S_z\rangle$ projections and the field dependence of $\omega_s/2\pi$ (in Hz) is better modeled by a second order polynomial $\Delta/2\pi = a_1(H_r-H_0)+a_2(H_r^2-H_0^2) = (H_r - H_0)\left[a_1 + a_2(H_r+H_0)\right]$ where $a_{1,2}$ are the first and second order polynomial coefficients respectively. Consequently, the fit functions used to extract $\gamma_s/2\pi$ and $g_c/2\pi$ (in hertz) are of the following form:
\begin{gather}
	\scriptstyle
	f_c' = f_c +m_{f_c}\left(H_r-H_0\right) + \frac{A\left(H_r-H_0\right)\left[a_2\left(H_r+H_0\right)+a_1\right]} {\left(H_r-H_0\right)^2\left[a_2\left(H_r+H_0\right)+a_1\right]^2+\left(\frac{\gamma_s}{2\pi}\right)^2}, \label{omegacfit} \\
	\scriptstyle
	\frac{\kappa_c'}{2\pi}=\frac{\kappa_0}{2\pi} +m_{\kappa_c}\left(H_r-H_0\right) + \frac{A \frac{\gamma_s}{2\pi}} {\left(H_r-H_0\right)^2\left[a_2\left(H_r+H_0\right)+a_1\right]^2+\left(\frac{\gamma_s}{2\pi}\right)^2}, 
	\label{kappacfit}
\end{gather}
where the first two terms approximated a linear background for the superconducting cavity field dependence and $A=\left(g_c/2\pi\right)^2$. An example of the fit performed is shown on Figs. 3(c-d) of the main article for a single resonance \textit{R}$_2^{-}$ and for all six resonances in the Fig.~\ref{figS2} where the linear background is removed for the frequency shift data. It is notable that the $\delta f_c$ perturbation extends over the entire field region around a resonance, whereas the peak of $\kappa_c'$ perturbation is located in a narrower region around a given resonance. The $\omega_s/2\pi$ as a function of field for each of the six transitions is shown in Fig.~\ref{figS3} (open diamonds); with a red line a linear fit is given ($R\approx1$) and second order polynomial fit (blue) gives $R=1$ within the precision of the fit software (Origin Labs). 

\begin{table*}[t]
	\caption{Eigenstates of the spin Hamiltonian for at the observed resonance fields for the transition $R_2^-$. Amplitude coefficients less than 0.01 have been disregarded for clarity.}
	\begin{tabular*}{\textwidth}{ccccccccc}
		\hline\hline
		\\
		\hspace{2mm}$|\psi_3\rangle = $\hspace{6mm}&\hspace{1mm} $S_z = 7/2$ \hspace{6mm}&\hspace{1mm} $S_z = 5/2$ \hspace{6mm}&\hspace{1mm} $S_z = 3/2$ \hspace{6mm}&\hspace{1mm} $S_z = 1/2$ \hspace{6mm}& \hspace{1mm}$S_z = -1/2$\hspace{6mm} & \hspace{1mm}$S_z = -3/2$\hspace{6mm} & \hspace{1mm}$S_z=-5/2$\hspace{6mm} & \hspace{1mm}$S_z=-7/2$\hspace{2mm} \\
		\hline
		$I_z = 3/2$ & 0 & 0 & 0 & 0 & 0 & 0 & 0 & 0 \\
		$I_z = 1/2$ & 0 & 0 & 0 & 0 & 0 & 0 & 0 & 0.0236 \\
		$I_z = -1/2$ & 0 & 0 & 0 & -0.0152 & 0 & 0 & 0.0317 & -0.9989 \\
		$I_z = -3/2$ & 0 & 0 & 0 & 0 & 0 & 0 & 0 & -0.0204 \\
		\hline
		\\
		\hspace{1mm}$|\psi_{14}\rangle = $\hspace{1mm}&\hspace{1mm} $S_z= 7/2$ \hspace{1mm}&\hspace{1mm} $S_z = 5/2$ \hspace{1mm}&\hspace{1mm} $S_z = 3/2$ \hspace{1mm}&\hspace{1mm} $S_z = 1/2$ \hspace{1mm}& \hspace{1mm}$S_z = -1/2$\hspace{1mm} & \hspace{1mm}$S_z = -3/2$\hspace{1mm} & \hspace{1mm}$S_z -5/2$\hspace{1mm} & \hspace{1mm}$S_z = -7/2$\hspace{1mm} \\
		\hline
		$I_z = 3/2$ & 0 & 0 & 0 & 0 & 0 & 0 & 0 & 0 \\
		$I_z = 1/2$ & 0 & -0.0226 & 0 & 0 & 0 & 0 & 0.0107 & 0 \\
		$I_z = -1/2$ & 0.0267 & 0.8376 & 0.0220 & 0 & 0 & 0.0215 & 0.5415 & 0.0171 \\
		$I_z = -3/2$ & 0 & 0.0425 & 0 & 0 & 0 & 0 & 0.0264 & 0 \\
	\end{tabular*}
	
	\label{tableS1}
\end{table*}

\section{Gd:CaWO$_4$ spin system and resonator coupling strength estimation} \label{sec2}

The $^{155,157}$Gd$^{3+}$:CaWO$_4$ spin Hamiltonian under the presence of a static magnetic field is described by Eq.(2) of the main article. Since $S=7/2$ and $I=3/2$, eigenstates $|\psi_{k=1\ldots32}\rangle$ are an admixture of basis states $|S_{z},I_{z}\rangle$. For a field parallel to the $c$-axis, the eigenstates are pure $|S_{z},I_{z}\rangle$ states and Electron Spin Resonance (ESR) transitions would fulfill the condition $\Delta S_z=1$ to be allowed. In perpendicular field configuration, as done in our experiment, transitions with large  $\Delta S_z$ are no longer forbidden due to the state admixtures. To demonstrate this aspect, we show in Table~\ref{tableS1} the composition of the two eigenstates, calculated with the EasySpin package, for the transition $R_2^-$ as an example. While the initial state $|\psi_3\rangle$ can be approximated by the pure state $|-7/2,-1/2\rangle$, the final state $|\psi_{14}\rangle$ is well approximated by a superposition of the states $|5/2, -1/2\rangle$ and $|-5/2,-1/2\rangle$ leading to a large $\Delta \langle S_z\rangle=4.52$. The ESR resonance has a transition element $\langle\psi_3|S_{x,y}|\psi_{14}\rangle\approx0.7$ meaning that for a microwave field perpendicular to the $c$~axis this resonance is no longer forbidden. A summary of the $\Delta \langle S_z\rangle$ values and the corresponding matrix transition elements for the observed resonances is given in Table~\ref{tableS2}.

Having a well known description of the eigenstates of the spin Hamiltonian we proceed to estimate their coupling to the electromagnetic field $\mathbf{H_1}$ of the cavity. The cavity power corresponding to vacuum fluctuations is given by $P_{n=0}=hf_c\kappa_c/2$, where $\frac{\kappa_c}{2\pi} = 5.84$~MHz. It is important to note that the higher the $f_c$, the higher the power and thus the corresponding vacuum rms field $(\delta\mathbf{H_1}^{\text{rms}})^2$ and spin-photon coupling strength; it is a characteristic of our experiment which is uses a relatively large frequency, close to 20~GHz.  For given initial $|\psi_i\rangle$ and final $|\psi_f\rangle$ states in resonance, the single spin coupling strength is given by:
\begin{equation}
	g_0(\mathbf{r}) = \gamma_e|\langle \psi_f | \delta\mathbf{H_1}^{\text{rms}}(\mathbf{r})\cdot \mathbf{S} | \psi_i \rangle|
\end{equation}
 where $\delta\mathbf{H_1}^{\text{rms}}(\mathbf{r})$ is the spatial distribution of the vacuum fluctuations obtained from the COMSOL finite-element calculations at $P_{in} = hf_c\kappa_c/2 =$~-126.7~dBm. The spin-ensemble coupling strength can therefore be estimated as~\cite{Kubo2010a, leDantec2021}:
 \begin{gather}
 	g_c = \sqrt{\rho\frac{\kappa_c}{\gamma_s}\frac{0.304}{2I+1}\int_{V_m}^{}d^3\mathbf{r}\left|g_0(\mathbf{r})\right|^2}\label{gens}
 \end{gather}
where the spin concentration $\rho$ is corrected by $min(\frac{\kappa_c}{\gamma_s},1)$ (spin selectivity as done by the cavity) and for the abundance of the $^{155,157}$Gd isotopes. Given the geometry of our resonator structure, the microwave field has components parallel and perpendicular to the crystallographic $c$-axis and thus two transitions can be excited with intensities $\propto\left|\langle \psi_f|S_{x,z}|\psi_i \rangle\right|^2$ respectively. Table~\ref{tableS2} presents a summary of the matrix transition elements $\langle \psi_f|S_{x,z}|\psi_i \rangle$ for the observed six spin resonances, using the Easyspin built-in \textit{sop} function~\cite{Stoll2006}. Since the observed resonances are allowed nuclear transitions ($\Delta I_z = 0$) the spin operators are defined as $S_{x,z}\cdot\mathbf{1_I}$ where $\mathbf{1_I}$ is the identity operator acting on the nuclear subspace. As seen on table~\ref{tableS2}, the $S_x$ transitions have the leading contribution and therefore the $g_c$ spin-resonator coupling strength can be estimated as:
\begin{equation}
	g_c = \gamma_e\left|\langle \psi_f|S_{x}|\psi_i \rangle\right| \sqrt{\rho\frac{\kappa_c}{\gamma_s}\frac{0.304}{4}\int_{V_m}^{}d^3\mathbf{r}\left|\delta H_{1,x}^{\text{rms}}(\mathbf{r})\right|^2}.
\end{equation}

The value of $g_c$ is then evaluated for increasing integration volumes $V_m$ by using different $H_{th}/H_1^{max}$ thresholds as seen in Fig(2)c of the main article. For a resonator mode volume of ~1100 $\mu$m$^3$ we found $g_c/2\pi$ in the order of 2-5MHz which agrees with the experimentally obtained values leading to $7.6\times 10^7$ spins excited per transition.
\\
\begin{table*}[t]
	\caption{Summary of the matrix transition elements and change of $\langle S_z\rangle$ values for the observed transitions.}
	\begin{tabular*}{0.6\textwidth}{cccc}
		\hline\hline
		\\
		\hspace{1mm}Transition\hspace{1mm}&\hspace{5mm}$\left|\langle \psi_f | S_x | \psi_i \rangle\right| \approx \left|\langle \psi_f | S_y | \psi_i \rangle\right|$ \hspace{1mm}&\hspace{5mm}$\left|\langle \psi_f | S_z | \psi_i \rangle\right|$\hspace{1mm}&\hspace{5mm} $ \Delta \langle S_z\rangle$\hspace{1mm}\\
		\hline
		$R_1^{-}$ (4 to 13) & 0.796 & 0.0255 & 4.14 \\
		$R_1^{+}$ (4 to 13) & 0.808 & 0.0245 & 4.085 \\
		$R_2^{-}$ (3 to 14) & 0.710 & 0.0165 & 4.52 \\
		$R_2^{+}$ (3 to 14) & 0.715 & 0.016 & 4.50 \\
		$R_3^{-}$ (2 to 15) & 0.546 & 0.008 & 5.12 \\
		$R_3^{+}$ (2 to 15) & 0.539 & 0.007 & 5.14 \\
		\hline
		
	\end{tabular*}
	\label{tableS2}
\end{table*}

\bibliography{cpsresref.bib}